\documentclass[runningheads]{llncs}
\usepackage[T1]{fontenc}
\usepackage{graphicx}

\usepackage{cite,hyperref,amsmath,graphicx,booktabs,amssymb,amsfonts,multirow,color,pifont}

\begin{document}
\title{Post-fusion Event-aware Time-Frequency Refinement for Audio Moment Retrieval}
\titlerunning{Post-fusion Event-aware Time-Frequency Refinement for AMR}
\author{
Feiyang Xiao\inst{1}$^{\dagger}$\ \and
Li'ang Luo\inst{1}$^{\dagger}$ \and
Kejia Zhang\inst{1} \and
Qiaoxi Zhu\inst{2} \and
Guangjun He\inst{3} \and
Pengming Feng\inst{3} \and
Jian Guan\inst{1}$^{\ast}$
}
\authorrunning{F. Xiao et al.}
\institute{%
College of Computer Science and Technology, Harbin Engineering University, Harbin, China \and
University of Technology Sydney, Ultimo, Australia \and
State Key Laboratory of Space Information System and Integrated Application, Beijing, China\\
\email{j.guan@hrbeu.edu.cn}
}

\maketitle

\begingroup
\renewcommand{\thefootnote}{$^{\dagger}$}
\renewcommand{\thefootnote}{$^{\ast}$}
\endgroup

\begin{abstract}

Audio Moment Retrieval (AMR) aims to locate the temporal segment within a long audio recording that semantically corresponds to a natural-language query. This task is particularly challenging because query-related audio events are often sparsely distributed and embedded in complex acoustic environments, where background sounds and acoustically similar events can easily obscure the target moments. Existing methods mainly strengthen audio-text semantic interaction to enhance the query-related information in audio frames, but seldom refine the resulting audio representations before temporal localization, limiting their ability to accurately distinguish target events from irrelevant acoustic interference. To address this issue, we propose ETR-AMR, a post-fusion event-aware time-frequency refinement framework for AMR. Specifically, an event-aware time-frequency refinement (ETR) strategy is proposed after audio-text interaction to refine fused audio frame features before temporal moment localization. ETR consists of two complementary mechanisms: time frame gating and frequency refinement. The time frame gating mechanism estimates frame-wise event relevance to emphasize query-related audio frames and suppress redundant temporal responses, while frequency refinement applies learnable time-frequency reweighting to enhance discriminative event-related frequency patterns and reduce background-dominant interference. By jointly refining fused audio frame features in the temporal and frequency domains, ETR produces more discriminative event-aware features for accurate temporal boundary prediction. Experiments on four AMR datasets demonstrate that ETR-AMR achieves superior performance compared with existing methods, especially on the high-quality CASTELLA benchmark. Duration-wise analysis further confirms the effectiveness of ETR for medium and long audio moments.   

\end{abstract}

\section{Introduction}
Audio Moment Retrieval (AMR) is an emerging multimodal understanding task that aims to localize query-related temporal segments within long audio recordings according to natural-language queries \cite{munakata2025audiomoment}. Unlike conventional sound event detection \cite{schmid2025effective,cai2025prototype}, which recognizes predefined event categories, AMR requires models to retrieve audio moments described by free-form textual queries with diverse semantic descriptions \cite{munakata2025audiomoment,munakata2026castella,wang2025qg}. This capability enables users to efficiently search for specific events from lengthy recordings, benefiting a wide range of real-world applications, including sports highlight retrieval \cite{sattari2025semantic}, surveillance event analysis \cite{munakata2025audiomoment}, and podcast navigation \cite{alexander2021audio}. Recently, the release of the CASTELLA benchmark \cite{munakata2026castella} has established the first large-scale human-annotated dataset for AMR, facilitating the development of DEtection TRansformer (DETR)-based retrieval frameworks \cite{moon2023query,carion2020end} that combine pretrained audio-language representations \cite{wu2023large} with temporal localization networks. Despite these encouraging advances, AMR remains a relatively underexplored task compared with video moment retrieval \cite{moon2023query,xiao2024bridging}, leaving considerable room for improving both cross-modal semantic understanding and temporal localization accuracy.

A fundamental challenge of AMR lies in accurately identifying text query-relevant audio moments from long-duration audio recordings \cite{munakata2025audiomoment,Xiao2024a,xiao2026meclapscore}. In practical scenarios, query-related audio moments are often sparsely distributed across audio recordings lasting several minutes. For example, audio recordings in the CASTELLA dataset may extend up to 360 seconds while containing only one or several short target audio moments \cite{munakata2026castella}. Consequently, audio moment retrieval models are required to jointly capture long-range temporal dependencies and precisely distinguish relevant acoustic events from abundant background sounds, silence, and unrelated acoustic activities \cite{munakata2025audiomoment}. This difficulty is further aggravated by the diversity of natural-language queries, which may describe complex acoustic scenes, multiple simultaneous sound sources, or subtle semantic relationships between events \cite{wang2025qg}. Moreover, temporal boundaries are often ambiguous because acoustic events exhibit varying durations and gradual transitions rather than clear start and end points \cite{yoshinaga2025training}. These characteristics make precise temporal localization of query-related audio moments in long audio recordings significantly more challenging than conventional clip-level audio-text retrieval \cite{xiao2023enhancing} and sound event detection \cite{barahona2024enhancing,fan2025graph}, requiring models to learn fine-grained semantic alignment while producing discriminative temporal representations for accurate localization.

Existing AMR methods primarily build upon DETR-based temporal localization frameworks originally developed for video moment retrieval \cite{munakata2025audiomoment,munakata2026castella,moon2023query}. Representative approaches, including AM-DETR \cite{munakata2025audiomoment}, UVCOM \cite{xiao2024bridging}, and QG-DETR \cite{wang2025qg}, first employ pretrained audio-language models, such as Contrastive Language-Audio Pretraining (CLAP) \cite{wu2023large}, to extract frame-level audio representations from long audio recordings and semantic representations from text queries. The extracted audio features and text features are then processed by transformer-based encoders and decoders to predict the temporal boundaries of query-related audio moments. These studies enhanced cross-modal interaction through query-guided cross-attention mechanisms \cite{sun2024tr,duan2023cross}, enabling textual queries to better guide audio feature learning. Although these methods have substantially improved audio moment retrieval performance, their improvements mainly focus on strengthening cross-modal feature fusion \cite{wang2025qg}. After audio-text interaction, the resulting fused audio frame features are typically forwarded directly to the localization decoder without further refinement. Consequently, query-irrelevant background sounds, environmental noises, and acoustically similar but semantically unrelated audio events remain embedded within the fused audio frame features, reducing the discriminability of query-related audio events. Such redundant responses become particularly problematic in long audio recordings, where irrelevant acoustic patterns accumulate over time and may dominate the learned fused audio frame features. As a result, AMR models are more likely to generate inaccurate temporal boundaries or attend to non-target regions, ultimately limiting audio moment retrieval performance.

To address the above limitation, we propose ETR-AMR, a post-fusion event-aware time-frequency refinement method for audio moment retrieval. The core idea is to shift the focus from merely improving audio-text interaction to refining the fused audio frame features before temporal moment localization. This design aims to produce more discriminative event-aware audio frame features, so that the AMR model can focus on query-related audio events rather than background sounds or irrelevant acoustic patterns. Specifically, the proposed ETR-AMR introduces an event-aware time-frequency refinement (ETR) strategy that combines two complementary mechanisms: time frame gating and frequency refinement. The time frame gating mechanism estimates frame-wise event relevance to emphasize query-related audio frames and suppress redundant temporal responses, while frequency refinement performs learnable time-frequency reweighting to enhance event-related frequency patterns and reduce background-dominant interference. By jointly refining fused audio frame features in both temporal and frequency domains, ETR-AMR provides more reliable inputs for accurate temporal boundary prediction. Since the proposed ETR strategy is independent of specific cross-modal interaction designs, it can be integrated into existing DETR-based AMR frameworks with negligible computational overhead. Extensive experiments on four existing audio moment retrieval datasets demonstrate that ETR-AMR consistently improves retrieval performance, confirming that post-fusion event-aware refinement is an effective complementary paradigm for long-duration audio moment retrieval.

\section{Proposed Method}

\begin{figure}
    \vspace{-5mm}
    \centering
    \includegraphics[width=0.96\linewidth]{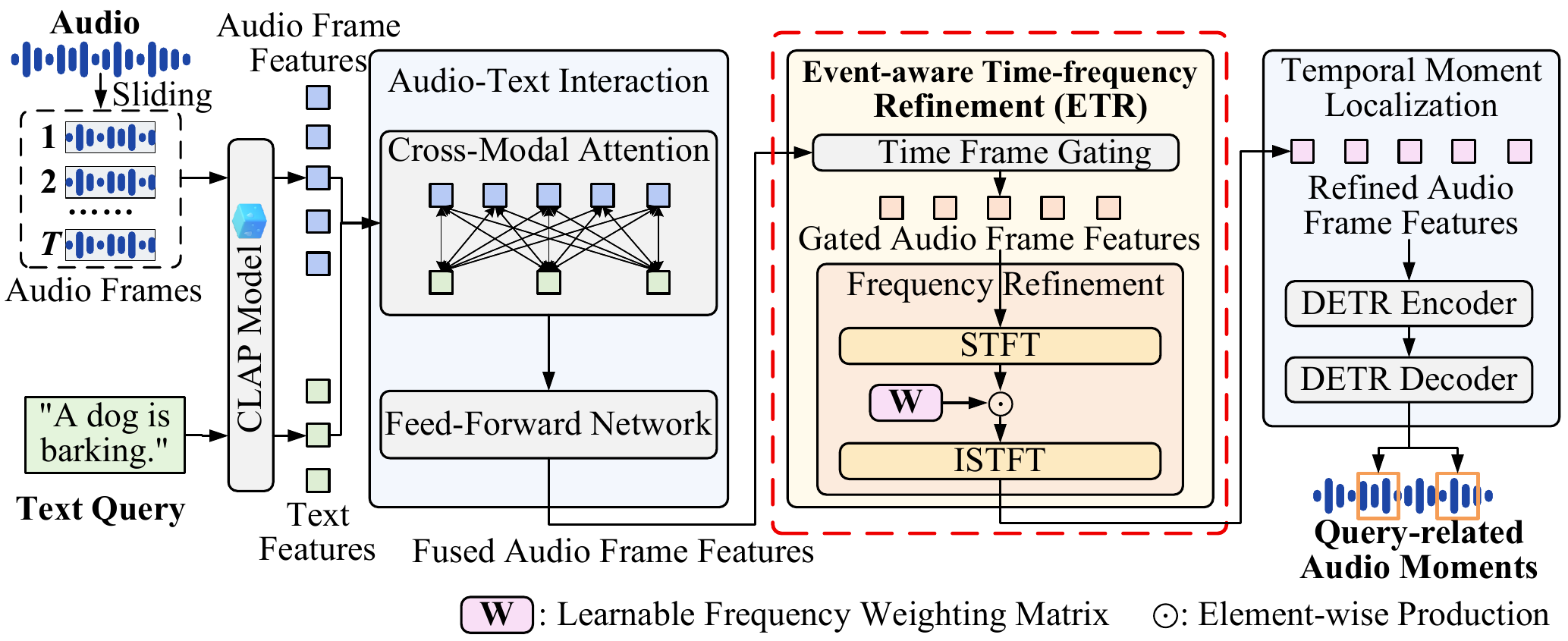}
    \vspace{-3mm}
    \caption{The overall framework of the proposed ETR-AMR method. ETR-AMR consists of three important stages: (1) audio-text interaction for fusing audio and text features, (2) the proposed event-aware time-frequency refinement to generate discriminative event-aware audio frame features through time-frequency refinement of audio features through the short-time Fourier transform (STFT), and (3) temporal moment localization to predict the start and end time of audio moments.}
    \label{fig:method}
    \vspace{-8mm}
\end{figure}

The proposed event-aware time-frequency refinement audio moment retrieval (ETR-AMR) is a post-fusion refinement framework for audio moment retrieval. Unlike existing AMR methods that mainly focus on audio-text interaction, ETR-AMR further refines the fused audio frame features before temporal moment localization, aiming to generate more discriminative event-aware representations for accurate boundary prediction. As illustrated in Figure~\ref{fig:method}, ETR-AMR consists of three stages: audio-text interaction, event-aware time-frequency refinement, and temporal moment localization. After cross-modal interaction, the fused audio frame features are refined through two complementary mechanisms. Time frame gating estimates frame-wise event relevance to enhance query-related audio frames and suppress redundant temporal responses, while frequency refinement applies learnable time-frequency reweighting to strengthen event-related frequency patterns and reduce background-dominant interference. Finally, the refined audio frame features are fed into a DETR-based localization module to predict query-related audio moments.

\subsection{Audio-Text Interaction}

The proposed ETR-AMR first performs cross-modal interaction between the input audio recording and the text query to obtain query-aware fused audio frame features. As illustrated in Figure~\ref{fig:method}, a long audio recording is first divided into a sequence of overlapping audio frames using a sliding-window strategy. Let $\mathcal{A}=[a_1,a_2,\ldots,a_T]$ denote the resulting audio frame sequence, where $T$ is the number of audio frames. Meanwhile, the input text query is represented as a sequence of word tokens, denoted by $\mathcal{Q}=[q_1,q_2,\ldots,q_L]$, where $L$ denotes the number of textual tokens.

A pretrained CLAP model \cite{wu2023large} is employed to encode both modalities into a shared embedding space. Specifically, each audio frame is independently encoded by the CLAP audio encoder to generate frame-level audio representations
\begin{equation}
    \mathbf{A}= \mathcal{F}_{\text{a}}( \mathcal{A} ) = [\mathbf{a}_1,\mathbf{a}_2,\ldots,\mathbf{a}_T]\in\mathbb{R}^{T\times C},
\end{equation}
where $C$ denotes the feature dimension, $\mathcal{F}_{\text{a}}(\cdot)$ denotes the audio encoder of the pretrained CLAP model. Similarly, the text query is encoded by the CLAP text encoder $\mathcal{F}_{\text{t}}(\cdot)$ to obtain textual representations
\begin{equation}
    \mathbf{Q}= \mathcal{F}_{\text{t}}( \mathcal{Q} ) = [\mathbf{q}_1,\mathbf{q}_2,\ldots,\mathbf{q}_L]\in\mathbb{R}^{L\times C}.
\end{equation}

Based on the extracted audio frame features and text features, following QG-DETR \cite{wang2025qg}, a query-guided cross-modal interaction is performed to establish semantic correspondence between the audio recording and the text query. During this process, each audio frame interacts with the entire text query to estimate their semantic relevance, enabling query-related audio frames to be enhanced while reducing the influence of query-irrelevant acoustic content. The resulting query-aware audio representations are subsequently processed by a feed-forward network (FFN) to further model feature interactions, yielding the fused audio frame features $\mathbf{F}\in\mathbb{R}^{T\times C}$.

These fused audio frame features encode query-aware semantic information within the audio recording, providing informative features for temporal localization. However, because the FFN mainly performs feature transformation without explicitly refining temporal audio contexts, the resulting representations inevitably retain query-irrelevant background sounds and redundant acoustic patterns. Therefore, instead of directly forwarding $\mathbf{F}$ to the DETR-based localization network as in existing AMR methods \cite{munakata2025audiomoment,wang2025qg}, we propose the event-aware time-frequency refinement strategy to further produce more discriminative event-aware audio frame features for temporal moment localization.

\subsection{Event-aware Time-frequency Refinement}

Although audio-text interaction injects query semantics into audio frame features, the obtained fused audio frame features may still contain strong responses from background sounds, silence, and query-irrelevant acoustic events. This problem is particularly severe in long audio recordings, where target events are sparsely distributed and irrelevant acoustic patterns may dominate the fused features. Directly feeding these features into the localization module may cause the model to attend to non-target regions or predict inaccurate temporal boundaries. To address this limitation, we propose event-aware time-frequency refinement (ETR), a post-fusion refinement strategy that improves the discriminability of fused audio frame features from both temporal and frequency perspectives. Different from existing methods that mainly strengthen audio-text interaction \cite{munakata2025audiomoment,wang2025qg}, our ETR strategy focuses on refining the already fused audio frame features before temporal moment localization.

\subsubsection{Time Frame Gating}

As illustrated in Figure~\ref{fig:method}, given the fused audio frame features $\mathbf{F}\in\mathbb{R}^{T\times C}$ obtained from the audio-text interaction stage, ETR first performs time frame gating to estimate the event relevance of each audio frame. Different from simply generating gating weights from $\mathbf{F}$ alone, the proposed gating mechanism jointly considers a learnable global vector $\mathbf{g}$, the fused audio frame features $\mathbf{F}$, and the textual query features $\mathbf{Q}$. Here, $\mathbf{g}$ is used to represent the global semantic context of the entire audio recording, providing complementary global information for frame-wise relevance estimation.

Specifically, a lightweight gating module $\Phi(\cdot)$ generates an event-aware frame-wise importance map conditioned on $\mathbf{g}$, $\mathbf{F}$, and $\mathbf{Q}$, and the fused audio frame features are reweighted as
\begin{equation}
    \mathbf{F}_{g} = \sigma(\Phi(\mathbf{g}; \mathbf{F}; \mathbf{Q})) \odot \mathbf{F},
\end{equation}
where $\sigma(\cdot)$ denotes the GELU activation function and $\odot$ denotes element-wise multiplication. In this formulation, the gating weights are estimated by considering both global audio semantics and query-conditioned frame-level features. Therefore, audio frames that are more relevant to the text query and consistent with the global audio context are assigned stronger responses, while frames dominated by background sounds or irrelevant acoustic activities are suppressed. The resulting gated audio frame features $\mathbf{F}_{g}$ provide a cleaner and more event-aware temporal basis for subsequent frequency refinement.

\subsubsection{Frequency Refinement}

Time frame gating suppresses irrelevant responses along the temporal dimension, but it does not explicitly model the time-frequency variations of acoustic events. Since many audio events are characterized by distinctive frequency patterns, ETR further introduces frequency refinement to enhance event-related frequency components. Specifically, the gated audio frame features are transformed into the time-frequency domain using the short-time Fourier transform (STFT). A learnable frequency weighting matrix $\mathbf{W}$ is then applied to adaptively modulate different frequency components, and the refined audio frame features are reconstructed by inverse STFT:
\begin{equation}
    \mathbf{F}_{r} = \mathcal{F}_{\mathrm{ISTFT}} \left( \mathbf{W} \odot \mathcal{F}_{\mathrm{STFT}}(\mathbf{F}_{g}) \right),
\end{equation}
where $\mathcal{F}_{\mathrm{STFT}}(\cdot)$ and $\mathcal{F}_{\mathrm{ISTFT}}(\cdot)$ denote STFT and inverse STFT, respectively. The learnable weighting matrix $\mathbf{W}$ enables ETR to emphasize frequency patterns associated with query-related audio events information while suppressing background-dominant frequency components. In this way, ETR performs explicit time-frequency modulation rather than conventional audio feature transformation in AMR methods.

Overall, the proposed ETR provides a post-fusion event-aware refinement strategy for audio moment retrieval. In the proposed ETR strategy, the time frame gating identifies query-relevant temporal regions, while the frequency refinement enhances discriminative event-related frequency patterns within the gated features. These two mechanisms are complementary and jointly produce refined audio frame features $\mathbf{F}_{r}\in\mathbb{R}^{T\times C}$ with stronger acoustic event responses and weaker background interference, providing more reliable inputs for DETR-based temporal moment localization.

\subsection{Temporal Moment Localization}

After event-aware time-frequency refinement, the refined audio frame features are delivered to the DETR-based localization module to predict the temporal boundaries of query-related audio moments. In this stage, the refined audio frame features $\mathbf{F}_{r}\in\mathbb{R}^{T\times C}$ are first processed by a Transformer-based DETR encoder to further model global temporal dependencies among audio frames. The encoded representations are then fed into a Transformer-based DETR decoder together with a set of learnable moment anchors, which progressively decode the temporal information and generate a fixed number of temporal boundaries of candidate audio moments.

Each prediction consists of a temporal boundary and a confidence score, represented as
\begin{equation}
    \hat{\mathbf{m}}_i=(\hat{s}_i,\hat{e}_i,\hat{c}_i),
\end{equation}
where $\hat{s}_i$ and $\hat{e}_i$ denote the predicted start and end timestamps of the $i$-th audio moment, respectively, and $\hat{c}_i$ denotes its confidence score. During training, the predicted moments are matched with the ground-truth annotations using the Hungarian matching algorithm \cite{kuhn1955hungarian}, and the network is optimized in an end-to-end manner using the same objective function as AM-DETR. Specifically, the overall loss consists of a classification loss and a temporal localization loss, where the latter combines the $\mathcal{L}_1$ regression loss and the generalized intersection-over-union (GIoU) loss for boundary regression. Since the proposed ETR only refines the input fused audio frame features without modifying the DETR-based localization module, it can be seamlessly integrated into the original DETR-based optimization framework.

\section{Experiments and Results}
\subsection{Experimental Setup}
\subsubsection{Dataset}

To comprehensively evaluate the effectiveness and generalization capability of the proposed ETR-AMR, experiments are conducted on four publicly available audio moment retrieval datasets, namely Clotho-Moment \cite{munakata2025audiomoment}, UnAV-100 subset \cite{geng2023dense}, TUT-Sound Events 2017 \cite{mesaros2016tut}, and CASTELLA \cite{munakata2026castella}. These datasets differ in terms of audio duration, recording scenarios, query diversity, and annotation quality, providing comprehensive evaluation under various acoustic conditions.

\textbf{Clotho-Moment} \cite{munakata2025audiomoment} is the first benchmark specifically developed for audio moment retrieval. It is constructed based on the Clotho audio captioning dataset \cite{drossos2020clotho} by automatically aligning textual descriptions with corresponding temporal audio segments. The dataset contains diverse everyday acoustic events and serves as a representative benchmark for evaluating open-domain audio moment retrieval.

\textbf{UnAV-100 subset} \cite{geng2023dense} is derived from the UnAV-100 dataset and contains long-form audio recordings covering diverse real-world acoustic environments. It provides complex audio scenes with multiple sound sources and overlapping acoustic events, making it suitable for evaluating retrieval robustness under challenging acoustic conditions.

\textbf{TUT-Sound Events 2017} \cite{mesaros2016tut} is adapted from a sound event detection benchmark. It contains real-life recordings with multiple overlapping sound events and temporal annotations, providing an additional evaluation scenario for assessing temporal localization performance in complex acoustic environments.

\textbf{CASTELLA} \cite{munakata2026castella} is the first large-scale audio moment retrieval benchmark with carefully curated and manually annotated audio moments and natural-language queries. It contains long-duration recordings of up to 360 seconds together with high-quality temporal annotations, providing more reliable and accurate ground-truth moment boundaries. Owing to its superior annotation quality, richer semantic diversity, and more realistic retrieval scenarios, CASTELLA is currently regarded as one of the most challenging and representative benchmarks for audio moment retrieval.

In this work, CASTELLA \cite{munakata2026castella} is adopted as the primary evaluation benchmark. The performance comparison and ablation studies are mainly conducted on this dataset to ensure reliable and fair evaluation. The remaining datasets are used to assess the generalization capability of the proposed ETR-AMR framework across diverse acoustic conditions.

\subsubsection{Implementation  Details}

In our experiments, all models are trained from scratch under the same experimental protocol. To ensure a fair comparison, identical training settings are adopted for all compared methods, including AM-DETR \cite{munakata2025audiomoment}, QG-DETR \cite{wang2025qg}, UVCOM \cite{munakata2026castella}, and the proposed ETR-AMR. The batch size is set to 32, and the initial learning rate is $1\times10^{-4}$. Unless otherwise specified, the remaining optimization settings follow those of QG-DETR. The proposed event-aware time-frequency refinement module is the only additional component introduced into the QG-DETR framework.

\subsubsection{Evaluation Metric}
Following previous AMR studies \cite{munakata2025audiomoment,munakata2026castella,wang2025qg}, the retrieval performance is evaluated using Recall at IoU thresholds (R1@50\% and R1@70\%) together with mean Average Precision (mAP@50\%, mAP@70\%, and mAP-avg).

\subsection{Performance Comparison}
\begin{table*}[t]
    \centering
    \caption{Performance comparison between the proposed ETR-AMR method and existing AMR methods on four AMR datasets.}
    \resizebox{\textwidth}{!}{
    \begin{tabular}{llccccc}
        \toprule
        Dataset & Method & R1@50\% & R1@70\% & mAP@50\% & mAP@70\% & mAP-avg \\
        \midrule

        \multirow{4}{*}{Clotho-Moment \cite{munakata2025audiomoment}}
        & AM-DETR \cite{munakata2025audiomoment} & 87.50 & 81.86 & 91.39 & 82.51 & 76.66 \\
        & QG-DETR \cite{wang2025qg} & \underline{88.64} & \underline{83.2}7 & \textbf{93.61} & 83.42 & 77.35 \\
        & UVCOM \cite{munakata2026castella} & \textbf{89.17} & \textbf{84.19} & \underline{92.67} & \textbf{84.95} & \textbf{80.48} \\
        & \textbf{ETR-AMR (Ours)} & 87.89 & 82.58 & 91.70 & \underline{83.50} & \underline{78.71} \\

        \midrule
        \multirow{4}{*}{UnAV-100 subset \cite{geng2023dense}}
        & AM-DETR \cite{munakata2025audiomoment} & 61.00 & 39.00 & 67.08 & 36.42 & 40.39 \\
        & QG-DETR \cite{wang2025qg} & \underline{63.00} & 40.00 & \underline{68.24} & 36.59 & 40.87 \\
        & UVCOM \cite{munakata2026castella} & \textbf{65.00} & \underline{47.00} & \textbf{70.41} & \underline{43.25} & \underline{44.27} \\
        & \textbf{ETR-AMR (Ours)} & \underline{63.00} & \textbf{48.00} & 67.18 & \textbf{44.20} & \textbf{44.70} \\

        \midrule
        \multirow{4}{*}{TUT-Sound Events 2017 \cite{mesaros2016tut}}
        & AM-DETR \cite{munakata2025audiomoment} & 14.42 & \underline{5.77} & \underline{15.72} & \underline{4.38} & \underline{6.23} \\
        & QG-DETR \cite{wang2025qg} & \underline{15.02} & 5.31 & \textbf{16.28} & \textbf{4.60} & \textbf{6.46} \\
        & UVCOM \cite{munakata2026castella} & \textbf{16.35} & \textbf{6.73} & 14.65 & 3.90 & 5.59 \\
        & \textbf{ETR-AMR (Ours)} & 11.54 & 4.81 & 12.88 & 3.81 & 5.41 \\

        \midrule

        \multirow{4}{*}{CASTELLA \cite{munakata2026castella}}
        & AM-DETR \cite{munakata2025audiomoment} & 19.30 & 7.05 & 15.93 & 4.82 & 6.41 \\
        & QG-DETR \cite{wang2025qg} & 13.88 & 5.86 & 13.28 & 4.33 & 5.71 \\
        & UVCOM \cite{munakata2026castella} & \underline{22.35} & \underline{9.87} & \underline{20.63} & \underline{6.81} & \underline{8.78} \\
        & \textbf{ETR-AMR (Ours)} & \textbf{29.40} & \textbf{15.07} & \textbf{25.35} & \textbf{11.15} & \textbf{12.63} \\

        \bottomrule
    \end{tabular}
    }
    \vspace{-5mm}
    \label{tab:res}
\end{table*}

We compare ETR-AMR with two representative existing DETR-based AMR methods, including AM-DETR \cite{munakata2025audiomoment} and QG-DETR \cite{wang2025qg}. AM-DETR is the first DETR-based framework for audio moment retrieval, while QG-DETR further improves audio-text interaction by introducing query-guided cross-modal feature enhancement. Compared with these methods, ETR-AMR focuses on refining the fused audio frame features after audio-text interaction, aiming to enhance query-related event responses and suppress irrelevant acoustic interference before temporal moment localization. Table~\ref{tab:res} presents the performance comparison between the proposed ETR-AMR and existing AMR methods on four benchmark datasets. 

As shown in Table~\ref{tab:res}, ETR-AMR achieves the best performance on the primary CASTELLA benchmark across all metrics. Compared with AM-DETR, ETR-AMR improves R1@50\% from 19.30\% to 29.40\%, R1@70\% from 7.05\% to 15.07\%, mAP@50\% from 15.93\% to 25.35\%, mAP@70\% from 4.82\% to 11.15\%, and mAP-avg from 6.41\% to 12.63\%. It also outperforms QG-DETR and UVCOM across all evaluation metrics on CASTELLA, demonstrating the effectiveness of refining fused audio frame features for audio moment retrieval in realistic long-audio scenarios with high-quality manual annotations.

On the UnAV-100 subset, ETR-AMR achieves the best R1@70\%, mAP@70\%, and mAP-avg, improving mAP-avg from 44.27\% to 44.70\% compared with the strongest baseline, UVCOM. This indicates that ETR-AMR performs favorably under stricter localization criteria and can improve temporal boundary accuracy. On Clotho-Moment, ETR-AMR achieves competitive performance, obtaining the second-best mAP@70\% and mAP-avg among the compared methods. These results demonstrate that ETR-AMR maintains competitive retrieval performance across different datasets while showing clear advantages under stricter localization criteria on the UnAV-100 subset.

However, ETR-AMR performs worse than AM-DETR and QG-DETR on TUT-Sound Events 2017. One possible reason is that TUT-Sound Events 2017 is originally designed for sound event detection rather than audio moment retrieval, and its event annotations are often shorter and more class-oriented. In addition, the dataset contains a large number of overlapping sound events within the same audio segments, which may further complicate temporal structure modeling and negatively affect the ability of ETR-AMR to capture clear temporal boundaries. In such cases, the proposed temporal refinement may be less effective because the dataset emphasizes fine-grained event detection rather than natural-language-guided long-audio moment retrieval. 

Overall, these results demonstrate that the proposed ETR-AMR is especially beneficial for AMR benchmarks with long audio recordings and high-quality natural-language annotations, confirming the effectiveness of event-aware time-frequency refinement for accurate audio moment retrieval.

\subsection{Visualization Analysis}

\begin{figure}
    \vspace{-5mm}
    \centering
    \includegraphics[width=0.8\linewidth]{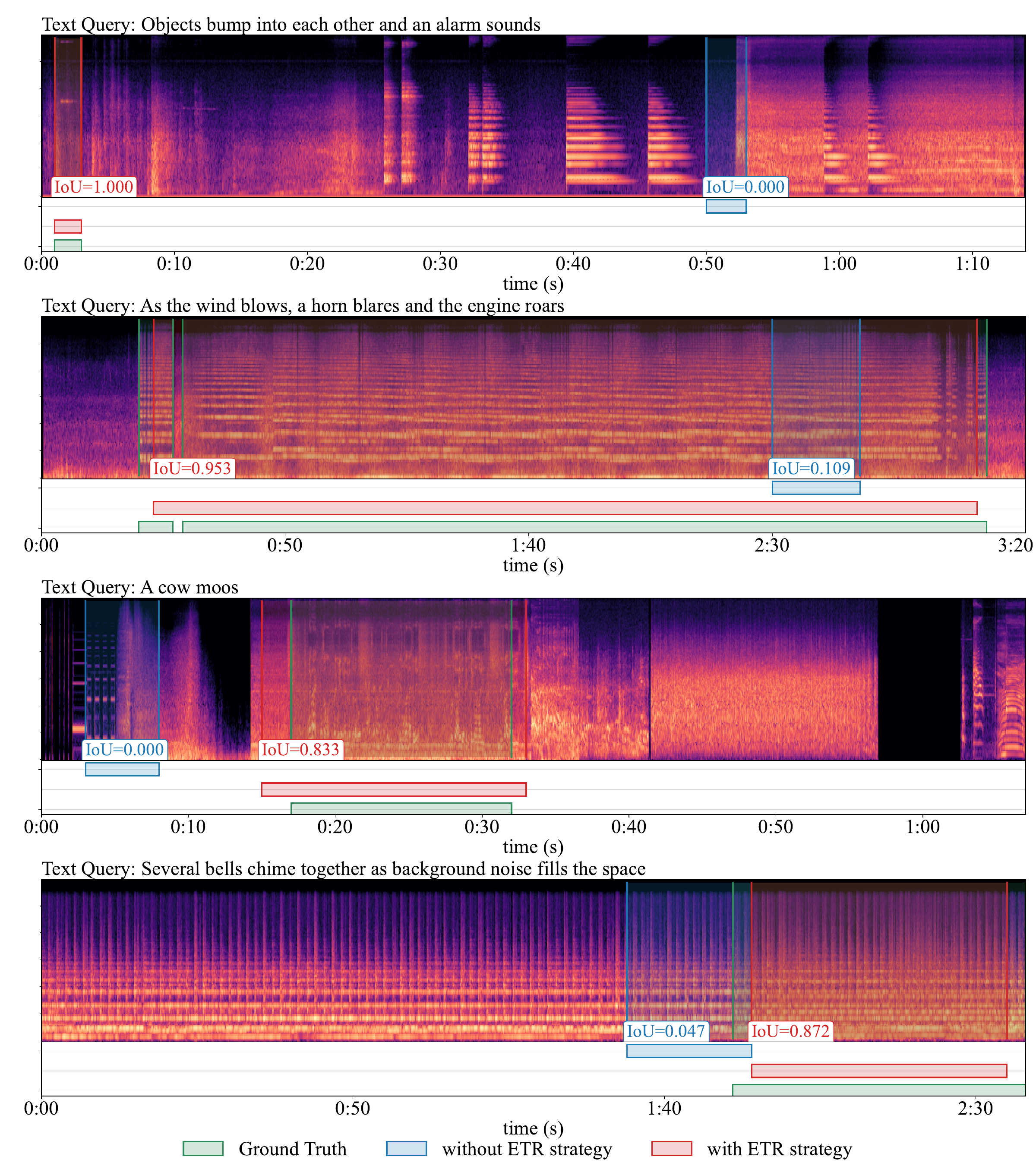}
    \vspace{-3mm}
    \caption{Visualization comparison between predictions without and with the proposed ETR strategy on the CASTELLA dataset \cite{munakata2026castella}. Green bars denote ground-truth moments, blue bars denote predictions without ETR, and red bars denote predictions with ETR.}
    \label{fig:vis}
    \vspace{-3mm}
\end{figure}

To further illustrate the effect of the proposed ETR strategy, we visualize several representative audio moment retrieval examples on the CASTELLA dataset \cite{munakata2026castella} in Figure~\ref{fig:vis}. The green bars denote the ground-truth audio moments, the blue bars denote the predictions without ETR, and the red bars denote the predictions with ETR. The corresponding IoU scores are also reported to directly compare the localization quality.

As shown in Figure~\ref{fig:vis}, the model without ETR often attends to query-irrelevant regions or produces temporal boundaries that deviate from the ground-truth moments. For example, in the first and third cases, the predictions without ETR are located in non-target regions, resulting in an IoU of 0.000. In contrast, the predictions with ETR are well aligned with the ground-truth moments, achieving substantially higher IoU scores. Similar improvements can also be observed in the second and fourth examples, where ETR significantly reduces boundary deviation and improves localization accuracy.

These visualization results provide intuitive evidence for the effectiveness of post-fusion event-aware refinement. By applying time frame gating and frequency refinement, ETR suppresses background-dominant and query-irrelevant responses while enhancing acoustic patterns related to the text query. Consequently, the localization decoder can focus on more discriminative audio frame features and predict temporal boundaries that better match the target audio moments.

\subsection{Ablation Study}

\subsubsection{Effectiveness on Different Moment Durations}

\begin{figure}
    \vspace{-5mm}
    \centering
    \includegraphics[width=0.99\linewidth]{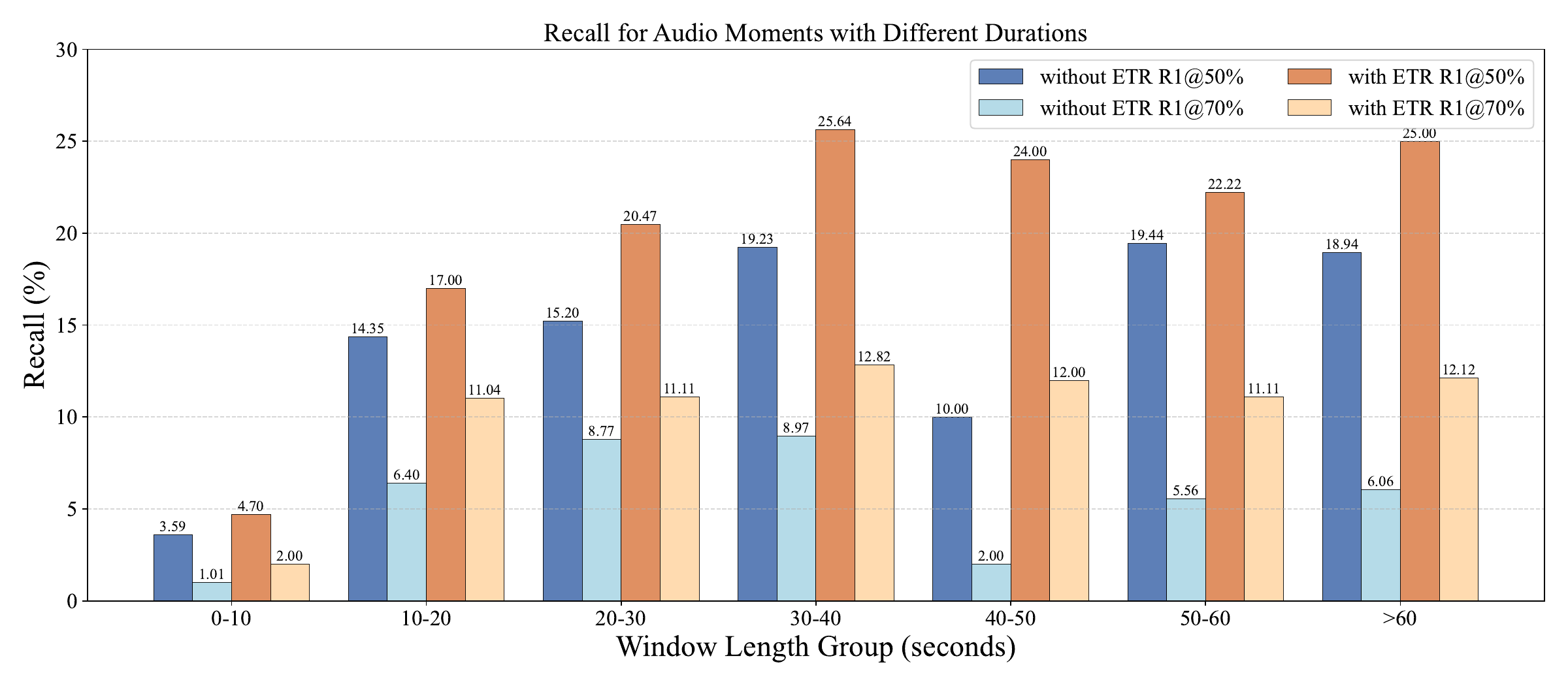}
    \vspace{-5mm}
    \caption{Duration-wise AMR performance comparison between the AMR model without ETR and the model with ETR on the CASTELLA dataset \cite{munakata2026castella}. R1@50\% and R1@70\% are reported for different audio moment duration groups.}
    \vspace{-5mm}
    \label{fig:duration}
\end{figure}

To explicitly evaluate the contribution of the proposed ETR strategy, we construct two variants of our ETR-AMR: one without ETR and the other with ETR. The former directly feeds the fused audio frame features into the localization module, while the latter inserts ETR after audio-text interaction to refine the fused audio frame features before temporal moment localization. This controlled comparison allows us to analyze whether ETR improves AMR performance under different audio moment durations. Specifically, audio moments in the CASTELLA dataset \cite{munakata2026castella} are grouped according to their temporal lengths, and R1@50\% and R1@70\% are computed for each duration group. The results are shown in Figure~\ref{fig:duration}.

As shown in Figure~\ref{fig:duration}, the model with ETR consistently outperforms the model without ETR across all duration groups under both R1@50\% and R1@70\%. For short moments within 0-10 seconds, both variants obtain relatively low recall, indicating that short audio events remain challenging due to limited temporal context and ambiguous boundaries. Nevertheless, introducing ETR improves R1@50\% from 3.59\% to 4.70\% and R1@70\% from 1.01\% to 2.00\%, showing that ETR can provide additional discriminative cues even for short-duration moments.

The improvements become more pronounced for medium and long audio moments. For example, in the 40-50 second group, adding ETR improves R1@50\% from 10.00\% to 24.00\% and R1@70\% from 2.00\% to 12.00\%. Similar gains are observed in the 50-60 second group and the longer-than-60-second group. These results suggest that ETR is particularly beneficial for longer audio moments, where background sounds and query-irrelevant acoustic patterns are more likely to accumulate within the fused audio frame features. By applying time frame gating and frequency refinement, ETR suppresses redundant acoustic responses and strengthens query-related event patterns, leading to more accurate temporal localization.

This duration-wise analysis verifies the effectiveness of the proposed ETR strategy. In particular, its advantage on medium and long audio moments demonstrates that post-fusion event-aware time-frequency refinement is useful for long-duration AMR scenarios, where temporal ambiguity and background interference are more severe.

\subsubsection{Effectiveness of Different Activations for Time Frame Gating}

\begin{table}[t]
    \centering
    \caption{Performance comparison on CASTELLA dataset \cite{munakata2026castella} when using different activations for the time frame gating in our proposed ETR module.}
    \begin{tabular}{lccccc}
        \toprule
        Activation & R1@50\% & R1@70\% & mAP@50\% & mAP@70\% & mAP-avg \\
        \midrule
        Sigmoid & 27.91 & 14.18 & \textbf{25.46} & 10.78 & 12.13 \\
        GELU & \textbf{29.40} & \textbf{15.07} & 25.35 & \textbf{11.15} & \textbf{12.63} \\
        \bottomrule
    \end{tabular}
    \vspace{-5mm}
    \label{tab:ablation_activation}
\end{table}

We further investigate the influence of different activation functions in the time frame gating mechanism. As shown in Table~\ref{tab:ablation_activation}, we compare Sigmoid and GELU as the activation $\sigma(\cdot)$ in the time frame gating mechanism on the CASTELLA dataset while keeping all other components unchanged. The activation function controls how the frame-wise gating responses are generated, thereby affecting the ability of ETR to emphasize query-related audio frames and suppress redundant temporal responses.

Compared with Sigmoid, GELU achieves better overall performance on most metrics. Specifically, GELU improves R1@50\% from 27.91\% to 29.40\%, R1@70\% from 14.18\% to 15.07\%, mAP@70\% from 10.78\% to 11.15\%, and mAP-avg from 12.13\% to 12.63\%. Although Sigmoid obtains a slightly higher mAP@50\%, GELU achieves better performance under the stricter IoU threshold and higher average mAP, indicating more accurate temporal boundary localization.

This improvement may be attributed to the smoother and non-saturating property of GELU, which allows the gating mechanism to preserve useful event-related responses rather than forcing frame weights into a highly compressed range. As a result, GELU provides more flexible frame-wise modulation for fused audio frame features, enabling ETR to better distinguish query-related events from background interference. Therefore, GELU is adopted as the activation function for time frame gating in our final ETR-AMR model.

\subsubsection{Effectiveness of Frequency Refinement}

\begin{table}[t]
    \centering
    \caption{Performance comparison on the CASTELLA dataset \cite{munakata2026castella} with and without frequency refinement in the proposed ETR module.}
    \begin{tabular}{cccccc}
        \toprule
        Frequency Refinement & R1@50\% & R1@70\% & mAP@50\% & mAP@70\% & mAP-avg \\
        \midrule
        \ding{55} & 27.32 & 12.32 & 24.29 & 9.90 & 11.52 \\
        \ding{51} & \textbf{29.40} & \textbf{15.07} & \textbf{25.35} & \textbf{11.15} & \textbf{12.63} \\
        \bottomrule
    \end{tabular}
    \label{tab:ablation_freq_refine}
\end{table}

To validate the contribution of frequency refinement in the proposed ETR module, we conduct an ablation study by removing this component while keeping the time frame gating mechanism unchanged. As shown in Table~\ref{tab:ablation_freq_refine}, the model without frequency refinement only performs temporal-domain gating on the fused audio frame features, while the full model further applies the frequency refinement by the learnable time-frequency reweighting after STFT.

The results show that introducing frequency refinement consistently improves all evaluation metrics on the CASTELLA dataset. Specifically, R1@50\% improves from 27.32\% to 29.40\%, R1@70\% improves from 12.32\% to 15.07\%, and mAP-avg improves from 11.52\% to 12.63\%. The larger gain on R1@70\% indicates that frequency refinement is particularly helpful for more accurate temporal boundary localization under stricter IoU criteria.

These improvements demonstrate that time-domain gating alone is insufficient to fully distinguish query-related audio events from background interference. By applying learnable reweighting in the time-frequency domain, frequency refinement enhances discriminative event-related frequency patterns and suppresses background-dominant components. Therefore, this component provides complementary refinement to time frame gating and contributes to more accurate audio moment retrieval.

\section{Conclusion}

In this paper, we proposed ETR-AMR, a post-fusion event-aware time-frequency refinement framework for audio moment retrieval. Unlike existing AMR methods that mainly focus on improving audio-text interaction, ETR-AMR introduces an event-aware time-frequency refinement module after cross-modal fusion to further refine fused audio frame features before temporal moment localization. By combining time frame gating and frequency-domain refinement, the proposed ETR module enhances query-related audio event responses while suppressing redundant background sounds and query-irrelevant acoustic patterns. Experiments on four AMR datasets demonstrate that ETR-AMR achieves superior performance compared with existing DETR-based methods, especially on the high-quality CASTELLA benchmark. Further duration-wise analysis shows that ETR is particularly effective for medium and long audio moments, where background interference and temporal ambiguity are more severe. These results confirm that post-fusion refinement provides an effective complementary strategy for improving accurate audio moment retrieval.

\subsubsection{Acknowledgments}
This work was partly supported by the Heilongjiang Provincial Natural Science Foundation of China under Grant No. BS2025F009, and the Greater Bay Area Institute of HPC-AI Co-Driven Innovation under Grant No. 03202604230020.

\subsubsection{Disclosure of Interests}
The authors have no competing interests to declare that are relevant to the content of this article.

%
%
\bibliographystyle{splncs04}
\bibliography{ref}

\end{document}